\documentclass[prd,twocolumn,showpacs,preprintnumbers,nofootinbib]{revtex4}
\usepackage{amsfonts,amsmath,amssymb}
\usepackage{graphicx}
\usepackage{color} 
\newcommand{\rthis}[1]{\textcolor{black}{#1}}

\usepackage[plainpages=false, colorlinks=true, anchorcolor=blue, linkcolor=blue, citecolor=blue, bookmarks=false]{hyperref}
\usepackage{natbib}
\pdfoutput=1

\usepackage{natbib}
\usepackage{breqn}

\makeatletter
\let\cat@comma@active\@empty
\makeatother

\begin{document}
\newcommand{\apjl}{Astrophys. J. Lett.}
\newcommand{\apjs}{Astrophys. J. Suppl. Ser.}
\newcommand{\aap}{Astron. \& Astrophys.}
\newcommand{\aj}{Astron. J.}
\newcommand{\araa}{Ann. Rev. Astron. Astrophys. } 
\newcommand{\mnras}{Mon. Not. R. Astron. Soc.}
\newcommand{\apss}{Astrophysics \& Space Sciences}
\newcommand{\jcap}{JCAP}
\newcommand{\pasj}{PASJ}
\newcommand{\pasp}{PASP}
\newcommand{\pasa}{Pub. Astro. Soc. Aust.}
\newcommand{\physrep}{Phys. Rep.}

\title{Bound on the graviton mass from   Chandra X-ray cluster sample}
\author{Sajal \surname{Gupta}$^1$}
\altaffiliation{E-mail: sg15ms084@iiserkol.ac.in}
\author{Shantanu  \surname{Desai}$^2$} \altaffiliation{E-mail: shntn05@gmail.com}

\affiliation{$^{1}$Department of Physical Sciences, IISER-Kolkata, Mohanpur, West Bengal-741246, India}

\affiliation{$^{2}$Department of Physics, Indian Institute of Technology, Hyderabad, Telangana-502285, India}
\begin{abstract}

We present new limits on the \rthis{graviton Compton wavelength in a Yukawa potential}  using  a sample of   12 relaxed galaxy clusters, for which   the temperature and gas density profiles were derived  by Vikhlinin et al~\citep{Vikhlinin06} using Chandra X-ray observations. \rthis{These limits can be converted to a bound on the graviton mass, assuming a non-zero graviton mass would  lead to a Yukawa potential at these scales.}
For this purpose, we first calculate the total dynamical mass from the hydrostatic equilibrium equation in  Yukawa gravity and then compare it with the corresponding mass in Newtonian gravity.  We calculate a 90 \% c.l. lower/upper limit on the graviton Compton wavelength/ mass for each of the 12 clusters in the sample. The best limit is obtained for Abell 2390, corresponding to     $\lambda_g > 3.58\times 10^{19}$ km or $m_g<3.46 \times 10^{-29}$ eV. \rthis{This is the first proof of principles demonstration of setting a limit on the graviton mass using a sample of related galaxy clusters with X-ray measurements and can be easily applied to upcoming X-ray surveys such as eRosita.}

\end{abstract}
\maketitle
\section{Introduction}
During the past decade there has been a resurgence of interest in massive gravity theories following breakthroughs with some of the long-standing vexing problems in these theories such as vDVZ discontinuity and Bouleware-Deser ghosts~\citep{Goldhaber10,derham11,derham14}.
On the observational/experimental side, there has been a renewed interest in obtaining improved limits on graviton mass from both astrophysical, laboratory and gravitational wave observations~\citep{Derham16,LIGO1}. 

Most recently, multiple groups have obtained such bounds on the graviton mass from galaxy clusters~\cite{Desai18,Rana,Gupta}, more than 40 years after the  first such limit with clusters~\citep{Goldhaber74}. Galaxy clusters are the most massive gravitationally collapsed objects in the universe. Exploiting the power of galaxy clusters for a wide variety of astrophysical (galaxy evolution), cosmological (dark energy, non-gaussianity), and fundamental physics (neutrino mass and modified gravity) studies is a key science goal for on-going and future dark energy surveys.  

The first ever limit on graviton mass from galaxy clusters ($m_g < 1.1 \times 10^{-29}$ eV)   was obtained, from the fact  that orbits of the largest known galaxy clusters at that time extended upto 0.58 Mpc~\cite{Goldhaber74}. However, some concerns about the assumptions made  to get this result have been pointed out~\cite{Desai18}. This result has been superseded by a more robust limit from Abell 1689,  corresponding  to $m_g< 1.37 \times 10^{-29}$ eV at \rthis{90\% c.l.}~\citep{Desai18}. Subsequently, Rana et al~\cite{Rana} obtained an updated limit of $5.9 \times 10^{-30}$ eV  \rthis{within $1\sigma$ confidence region} using the mass measurements from  Weak lensing and SZ -based galaxy cluster catalogs. This technique was then extended to other catalogs  and the best limit was obtained   using the SDSS redMaPPer catalog, given by  $m_g<1.27\times 10^{-30}$ eV~\citep{Gupta} \rthis{at 90\% c.l.} 

Here, we shall use the measured temperature and density profiles of 12 galaxy clusters~\citep{Vikhlinin06}, obtained using X-ray measurements with the Chandra X-ray satellite,  to get revised graviton mass bounds. The key idea in this work, similar to Refs.~\citep{Desai18,Rana} is to look for signatures of Yukawa gravity in these clusters. As discussed in Ref.~\cite{Derham16}, most of the massive gravity models give rise to a Yukawa potential in the non-relativistic decoupling limit of the theory. \rthis{Therefore, the  main presume for this work  (similar to previous works~\cite{Hare,Goldhaber74,Will97,Desai18,Rana,Gupta,Zakharov18} on bounding the graviton mass) is that a non-zero graviton mass would lead to a Yukawa potential on such scales}.

The outline of this paper is as follows. We do a brief recap of the Chandra X-ray data for the 12 clusters used in our analysis in Sect.~\ref{sec:1}. We describe the formalism  used for bounding the graviton mass  in Sect.~\ref{sec:analysis}.  We present our results in Sect.~\ref{sec:results} and conclude in Sect.~\ref{sec:conclusions}.

\section{Chandra X-ray cluster sample}
\label{sec:1}
Vikhlinin et al~\citep{Vikhlinin06} (V06, hereafter) presented density and temperature profiles for a  total of 13 nearby relaxed galaxy clusters (A133, A262, A383, A478, A907, A1413, A1795, A1991, A2029, A2390, MKW4, RXJ1159+55531, USGC 2152) using measurements  from the archival or pointed observations with the Chandra X-ray satellite. These measurements extended up to 1 Mpc for some of the clusters.  The typical exposure times ranged from 30-130 Ksecs. The temperatures span a range between  1 and 10 keV and masses from $(0.5-10) \times 10^{14} M_{\odot}$. In some cases, the Chandra data was supplemented with data from the ROSAT satellite to model the gas density. More details on these observations and their results can be found in V06 (see also Ref.~\cite{Vikhlinin05}). The main goal of this work was to reconstruct gas and total mass estimates, as well as the gas mass fraction for these galaxy clusters.
This  same data has previously been used to test multiple alternate gravity theories and non-standard dark matter scenarios~\citep{Rahvar,Khoury17,Bernal,Ng,Edmonds,Hodson17}.  We used 12 out of these 13 clusters (excluding USGC 2152, as some of the pertinent data was not available to us) in order to obtain a  limit on the graviton mass.

\section{Analysis}
\label{sec:analysis}
\subsection{Hydrostatic equilibrium masses}
To  get a bound on the graviton mass, we \rthis{first compute the dynamical mass in both Newtonian and Yukawa gravity (similar in spirit to the analysis in ~\cite{Rahvar}) and quantify the deviations between them}. For this purpose, we first consider the equation of hydrostatic equilibrium used for the mass determination in both Newtonian and Yukawa gravity.

\rthis{If we consider a gas in hydrostatic equilibrium, the pressure gradient $dP/dr$ balances the acceleration due to gravity $g(r)$, giving rise to $dP/dr=-\rho_g(r) g(r)$ where $\rho_g(r)$ is the mass density of the cluster gas at a distance $r$~\cite{Rahvar}. For Newtonian gravity $g(r)=\frac{G M(r)}{r^2}$, where $M(r)$ is the total dynamical mass at a distance $r$ from the cluster center. The gas pressure can be related to the density, assuming  an ideal gas equation of state $P=\rho K_b T/\mu m_p G$, where $m_p$ is the mass of the proton,  $\mu$ is the mean molecular weight of the cluster in a.m.u. and is approximately equal  to 0.6~\cite{Vikhlinin05,Rahvar}.  Putting all this together, the total dynamical mass for a spherically symmetric relaxed cluster  in hydrostatic equilibrium for an ideal gas equation of state
under the premise of Newtonian gravity ($M(r) \equiv M_{tot}^N(r)$) is  given by~\cite{Allen}:}
\begin{equation}
M_{tot}^N(r) = -\frac{k_bT r}{G\mu m_p}\left(\frac{d\ln\rho_{gas}}{d\ln r}+ \frac{d\ln T}{d\ln r}\right),
\label{eq:1}
\end{equation}
\rthis{For a non-zero graviton mass ($m_g$), the corresponding equation of hydrostatic equilibrium can be generalized by replacing the Newtonian acceleration ($g(r)$) with the  corresponding acceleration in a Yukawa potential ($a_{yuk}$)~\cite{Will97,Desai18,Rana}}
\begin{equation}
a_{yuk} = \frac{GM^{yuk}_{tot}(r)}{r} \exp \left(\frac{-r}{\lambda_g}\right) \left(\frac{1}{\lambda_g}+ \frac{1}{r}\right), 
\label{eq:accyuk}
\end{equation}
where $M^{yuk}_{tot}(r)$ is the total dynamical mass in Yukawa gravity;  $\lambda_g$ is the Compton wavelength of the graviton and is given by $\lambda_g \equiv \frac{h}{m_g c}$ for a graviton of mass $m_g$. We can calculate  $M^{yuk}_{tot}(r)$, by balancing the pressure gradient with the gravitational force felt by the gas of density $\rho_g(r)$, using $\frac{dP}{dr}=-\rho_g (r) a_{yuk}$. Therefore, plugging $a_{yuk}$ from 
Eq.~\ref{eq:accyuk} and assuming an ideal gas equation of state as before, we get the total dynamical mass in a  Yukawa potential
\begin{widetext}
\begin{equation}
M^{yuk}_{tot}(r)  = - \exp\left(r/\lambda_g\right)\frac{k_b T r}{G\mu m_p}\left(\frac{d\ln\rho_{gas}}{d\ln r}+ \frac{d\ln T}{d\ln r}\right)\frac{r \lambda_g}{\lambda_g+r},
\label{eq:yukhyd}
\end{equation}
\end{widetext}

We note that there are  alternate expressions for the   hydrostatic mass in  Yukawa gravity~\cite{Bertolami}. However in that work~\cite{Bertolami}, the Yukawa potential considered is different than the one considered here and cannot be used to estimate the graviton mass, as it does not reduce to the Newtonian potential in the limit that the graviton mass tends to zero.

\subsection{Temperature and density profiles}
The first step in calculating the mass profiles for both Yukawa and Newtonian gravity involves positing a model for the  gas and temperature profile as a function of distance from the cluster center. For this purpose, we  used the models from  V06, which were fit to the observed data. 

Let us first consider the gas profile. The hot plasma in galaxy clusters emits X-rays via thermal bremmsstrahlung.  The intensity of the X-ray emissions proportional to the number density of electrons ($n_e$) and protons ($n_p$). This product is related to the gas density~\citep{Vikhlinin05} 
\begin{equation}
\rho_g \approx 1.624 m_p\sqrt{n_p(r)n_e(r)},
\label{gasdensity}
\end{equation}
\noindent for a plasma with primordial helium abundance and with metallicity equal to  $0.2 Z_{\odot}$. 

\rthis{
 The most widely adopted functional form for the gas density  in galaxy clusters  is the $\beta$-profile~\cite{betamodel}, which was obtained from the equation of hydrostatic equilibrium
for an isothermal gas, and assuming that the  matter distribution is governed by the King's profile.
To fit the observed X-ray emission, various modifications were made by V06 to the original beta profile~\cite{betamodel}. V06 used a superposition of two $\beta$ profiles with separate scale factors. The extra components were added  to account for the steepening brightness at $r  \simeq 0.3r_{200}$ and to have a cusp at  the center.
The final parametric form posited  for $n_p(r)n_e(r)$ in V06, which best fits the X-ray   is given by} 
\begin{widetext}
\begin{equation}
n_e (r)n_p(r) = \frac{(r/r_c)^{-\alpha'}}{(1+r^2/r_c^2)^{3\beta -\alpha' /2}}\frac{n_0^2}{(1+r^\gamma/r_s^\gamma)^{\epsilon/\gamma}} + \frac{n_{02}^2}{(1+r^2/r_{c'}^{2})^{3\beta'}}.
\label{eq:gasdensity}
\end{equation}

\end{widetext}
The physical  interpretations of  the empirical constants $r_c$,$\alpha'$, $\beta$, $r_s$, $\gamma$, $n_0$, $n_{02}$, $\beta'$ for the twelve galaxy clusters are discussed in V06 and can be found in Table 2 therein, as well as in Table III of Ref.~\cite{Rahvar}. \rthis{ We note that although more physically motivated functional forms for the gas density profiles have been proposed~\cite{Komatsu,Loeb}, there is some degeneracy between these profiles and the associated theory of gravity,  cosmological model
as well as the dark matter density distribution. Since it is not possible to derive an ab initio model-independent estimate of the gas density profile,
usually some variant of $\beta$ profile is always used to parameterize the gas density in clusters. Previously, the  gas density profile from Eq.~\ref{eq:gasdensity} for the same sample has also been  used for  cosmological parameter estimation~\cite{Vikhlinin09}, tests of alternate gravity theories~\cite{Rahvar,Hodson17,Khoury17}, and also tests of alternate dark matter scenarios~\cite{Bernal,Edmonds,Ng}.  Therefore, we too use the same  profile for our work, since they fit the X-ray surface brightness observations. For calculating the limit on graviton mass, we only need the derivative of the logarithm of the density  profile, which we use from Ref.~\cite{Rahvar}. }

\rthis{The X-ray emission energy spectrum for all the 11 clusters was modeled using  the Mekal model~\cite{Vikhlinin05}. The temperature can then be directly estimated after positing a metallicity and gas density profile. These temperature measurements along with error bars can be found in Refs.~\cite{Vikhlinin05,Vikhlinin06}.}

The observed temperature profile in these galaxy clusters peaks at 0.1-0.2 $r_{200}$ and falls off thereafter.  It also shows a decline near the center of the cluster. 
In V06, an analytic model for the temperature profile was constructed to describe these gross observational features. The  3D temperature profile ($T(r)$) used for each of these clusters is given by ~\cite{Vikhlinin06,Allen01},
\begin{equation}
T(r) = T_0 \frac{(x_0+T_{min}/T_0)}{x_0+1}\frac{(r/r_t)^{-a'}}{\left[1+(r/r_t)^b\right]^{c'/b}},
\label{eq:temp}
\end{equation}
where $x_0=\left(\frac{r}{r_{cool}}\right)^{a_{cool}}$. The  physical meanings of the eight free  parameters $a',b,c',T_{min},r_t,T_0,r_{cool}$, and $a_{cool}$  and their corresponding values for the 12 clusters can be found in V06 or  Ref.~\cite{Rahvar}. The observed values of $T(r)$ along with their error bars  at various points from the cluster center were provided for 12 out of the 13 clusters by A. Vikhlinin (private communication). \rthis{To calculate the dynamical mass, we need the derivative of the logarithm of the temperature described in Eq~\ref{eq:temp}, which we use from Ref.~\cite{Rahvar}. Similar to the gas density profile, there is also a degeneracy between the temperature profile and the underlying cosmological, dark matter as well the theory of gravity
and it is not possible to obtain a completely theory-agnostic form for the temperature profile from first principles.
Therefore, similar to previous works~\citep{Rahvar,Hodson17,Khoury17,Ng,Edmonds,Bernal}, which have used this data for testing non-standard models, we use the same temperature profiles from V06.}

\subsection{$\chi^2$ Definition} 
To get the corresponding limit on the graviton mass, we compare the dynamical masses in Newtonian and Yukawa gravity. This assumes that the dynamical hydrostatic  mass estimate  using Newtonian gravity is the true mass. The hydrostatic mass estimate for clusters assuming a Newtonian potential agrees with Weak lensing mass estimates, once you correct for the hydrostatic bias~\cite{Comalit}. So the X-ray mass assuming  a Newtonian potential can be considered the "truth", against which alternate models can be compared. Note that  in Ref.~\cite{Rahvar} also, the dynamical masses for the non-local gravity model was compared against the Newtonian mass estimate to constrain the alternate model.

Therefore, we calculate the $\chi^2$  differences between Newtonian and Yukawa masses
\begin{equation}
\chi^2= \sum\limits_{i=1}^N \left(\frac{M^{yuk}_{tot}(r)-M^{N}_{tot}(r)} {\sigma_{M^{N}_{tot}}}\right)^2,
\label{eq:5}
\end{equation}
where $\sigma_{M^{N}_{tot}}$  is the error in  $M^{N}_{tot}$. For each cluster, $\chi^2$ was evaluated at these points for which the errors in temperature and radii were available, allowing us
to do error propagation.

To evaluate the error in the  mass, we need to add in quadrature, the  errors in temperature, and distance. We do not include the errors in density, since no errors were provided for the parameters governing the gas density profile. 
\begin{equation}
\label{eq:4}
\sigma_{M^{N}_{tot}} = \left[\left(\frac{\partial M^{N}_{tot}}{\partial T}\right)^2~\sigma_{T}^2 + \left(\frac{\partial M^{N}_{tot}}{\partial r}\right)^2~\sigma_{r}^2  \right]^{1/2},
\end{equation}
where $\sigma_T$ and $\sigma_r$ denote the errors in the measurement of temperature and radius. We used the errors in distance and temperature provided to us by A.Vikhlinin. The partial 
derivative  $\frac{\partial M^{N}_{tot}}{\partial T}$  was obtained from Eq~\ref{eq:temp}.
Once $\sigma_{M^{N}_{tot}}$ is calculated in this way, it can be directly plugged in Eq.~\ref{eq:5}.

\section{Results}
\label{sec:results}
To determine the upper limit on the graviton mass, we evaluated Eq.~\ref{eq:5}  separately for each of the twelve clusters, by finding the mass corresponding to $\Delta \chi^2=2.71$~\citep{NR,Messier},  to get a  90\% c.l. upper limit on the graviton mass or a lower limit on the graviton Compton wavelength. The $\chi^2$ differences as a function of graviton mass can be found in Fig.~1. The corresponding 90\%  c.l. upper limits be found in Table~\ref{tab:results}. All  mass limits are $\mathcal{O}$($10^{-28}-10^{-29}$) eV or $\mathcal{O}$($10^{19}$) km in terms of  $\lambda_g$.  The best limit we obtained is for Abell 2390 (A 2390), corresponding to $m_g < 3.46 \times 10^{-29}$ eV. The reason these limits are of the same order of magnitude is because the cluster sample is very homogeneous. The temperature and density profiles show the same qualitative trends and the measurements are of the same order of magnitude. In fact as discussed in V06, for $r \geq r_{500}$, the scaled three-dimensional temperatures for all the clusters are within $\pm 15$\% from the average profile. The main sensitivity to the graviton mass comes from  the maximum distance up to which we calculate the accelerations. For all the clusters, this distance is about 1 Mpc (cf. Figs 3-15 of V06). Therefore, the graviton mass limits are of the same order of magnitude  for all the clusters. The best limit comes from A 2390, which is the only cluster for which we use one observation beyond 1 Mpc.

Although these limits are not as stringent as those obtained by stacking the cluster catalogs~\citep{Rana,Gupta},  our limits  are obtained using an independent analysis method   and using single-cluster data.  This is also the first result obtained using only X-ray measurements, whereas the previous results~\citep{Desai18,Gupta,Rana} used optical and SZ data.
Furthermore, limits  herein are more model-independent than the previous bounds from clusters. In Refs.~\citep{Desai18,Gupta,Rana}, NFW profile has been used to model the dark matter density distribution, and this profile is valid only for Newtonian gravity~\citep{NFW}. Here, we have not used any dark matter profile to obtain our bounds. The temperature and gas density profiles (which have been  estimated from X-ray surface brightness measurements, caused mainly by thermal Bremsstrahlung emission) are inferred directly from the observational data. Therefore, these same measurements have also been used to test multiple alternate gravity theories~\cite{Rahvar,Khoury17,Hodson17,Edmonds}.
The main {\it ansatz} used here is that all the clusters are relaxed and one can apply the corresponding equation of hydrostatic equilibrium. The eRosita X-ray satellite~\citep{erosita} is expected to be launched in 2019 and the same analysis can be applied to eRosita measurements of galaxy clusters to get more stringent bounds.

\begin{figure*} 
\centering
\includegraphics[width=\textwidth]{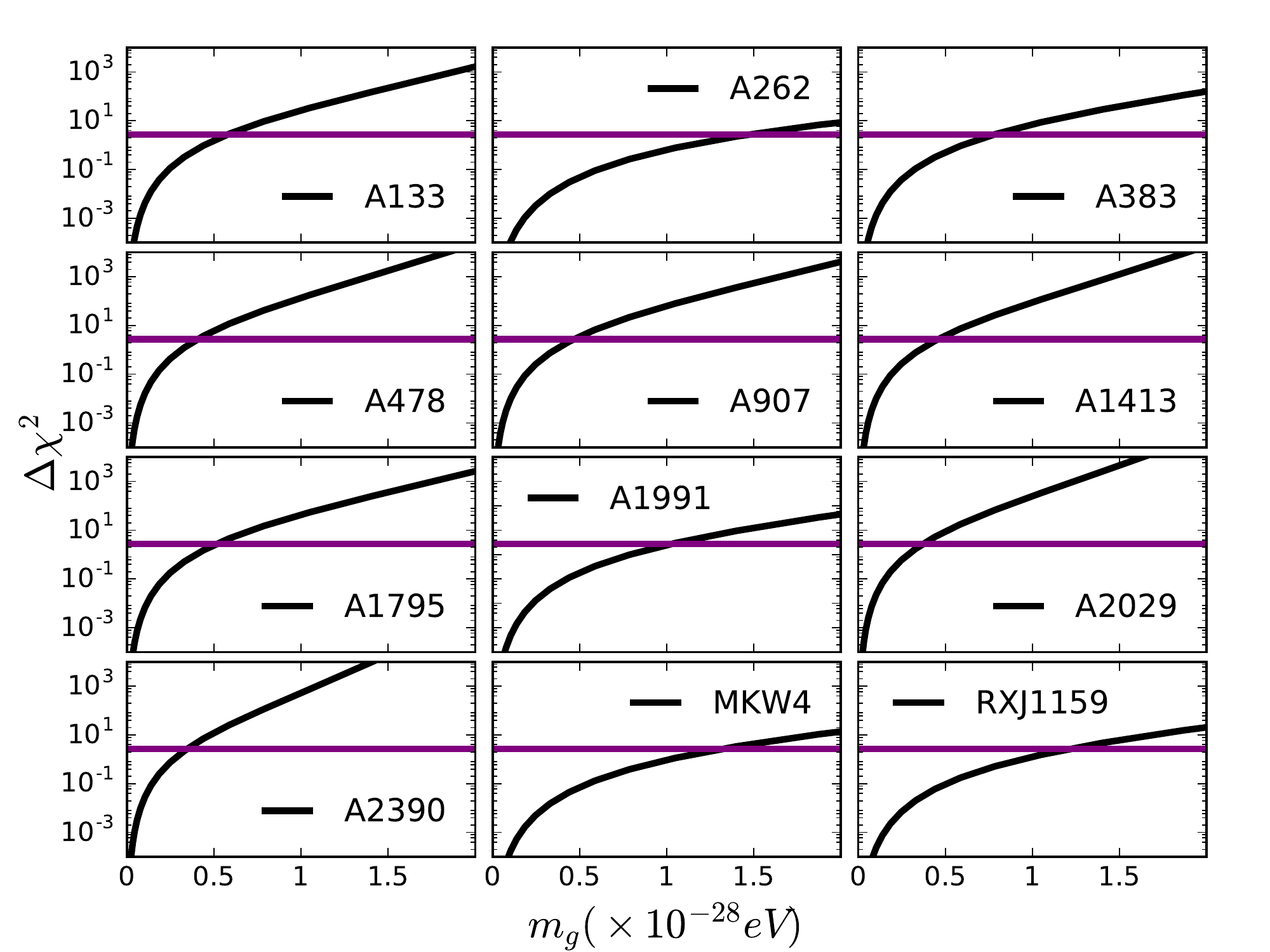}
\caption{$\Delta \chi^2$ as a function of graviton mass for each of the 12 clusters used for the analysis in each of the sub-panels. The magenta line in each sub-panel corresponds to the ordinate of 2.71, from which the corresponding 90\% c.l. on graviton mass can be determined. Summary of these limits for each cluster can be found in Tab.~\ref{tab:results}.}
\label{fig:2}
\end{figure*}


\begin{table}[t]
\vspace{5mm}
\begin{tabular}{|l|l|l|}
\hline
Cluster  Name & m$_g < $ (eV)            & $\lambda_g > $ (km)     \\ \hline
A 133         & 5.76 $\times$ 10$^{-29}$ ~~~~& 2.15 $\times$ 10$^{19}$ ~~~~\\ \hline
A 262          & 1.47 $\times$ 10$^{-28}$ ~~~~& 8.44 $\times$ 10$^{18}$ ~~~~\\ \hline
A 383          & 7.80 $\times$ 10$^{-29}$ ~~~~& 1.59 $\times$ 10$^{19}$ ~~~~\\ \hline
A 478          & 4.04 $\times$ 10$^{-29}$ ~~~~& 3.06 $\times$ 10$^{19}$ ~~~~\\ \hline
A 907          & 4.65 $\times$ 10$^{-29}$ ~~~~& 2.66 $\times$ 10$^{19}$ ~~~~\\ \hline
A 1413         & 4.57 $\times$ 10$^{-29}$ ~~~~& 2.71 $\times$ 10$^{19}$ ~~~~\\ \hline
A 1795         & 5.12 $\times$ 10$^{-29}$ ~~~~& 2.42 $\times$ 10$^{19}$ ~~~~\\ \hline
A 1991         & 1.02 $\times$ 10$^{-28}$ ~~~~& 1.21 $\times$ 10$^{19}$ ~~~~\\ \hline
A 2029         & 3.70 $\times$ 10$^{-29}$ ~~~~& 3.34 $\times$ 10$^{19}$ ~~~~\\ \hline
A 2390         & 3.46 $\times$ 10$^{-29}$ ~~~~& 3.58 $\times$ 10$^{19}$ ~~~~\\ \hline
MKW 4          & 1.32 $\times$ 10$^{-28}$ ~~~~& 9.38 $\times$ 10$^{18}$ ~~~~\\ \hline
RX J1159+5531           & 1.21 $\times$ 10$^{-28}$ ~~~~& 1.02 $\times$ 10$^{19}$ ~~~~\\ \hline
\end{tabular}
\caption{90\% confidence level upper (lower) limit on graviton mass (Compton wavelength) for each of the 12 galaxy clusters used in our analysis. The letter 'A' in the prefix of some of the clusters is an acronym for Abell. The best limit is for Abell 2390 or  A2390 ($m_g<3.46 \times 10^{-29}$ eV or $\lambda_g >3.58 \times 10^{19}$ km).}
\label{tab:results}
\end{table}


\section{Conclusions}
\label{sec:conclusions}

We have obtained lower limits on the \rthis{graviton  Compton wavelength for a Yukawa potential}, from the temperature and density profiles, obtained  by Vikhlinin et al~\citep{Vikhlinin05,Vikhlinin06}, for  a sample of 12 relaxed galaxy clusters using Chandra X-ray data. \rthis{Assuming a non-zero graviton mass gives rise to such a potential at the length scale of galaxy clusters, we inferred an upper limit on the graviton mass.}
From the equation for hydrostatic equilibrium for a massive graviton, we obtained the total dynamical mass in a Newtonian potential for each of these 12 clusters. \rthis{We then calculated the same for a Yukawa potential.} Then, we computed the $\chi^2$ deviations between these masses  as a function of radius. These differences can be found in Figure 1.  The limit on the graviton mass for each of these clusters was obtained,  from $\Delta \chi^2=2.71$. Our results can be found in Table~\ref{tab:results}. The best limit was obtained for  Abell 2390 (A2390), corresponding to $m_g<3.46 \times 10^{-29}$ eV, or $\lambda_g >3.58 \times 10^{19}$ km. Although, this limit is of the same order of magnitude as some the previous existing limits on graviton mass using clusters~\cite{Goldhaber74,Desai18} and is almost two orders of magnitude less stringent than the current best bounds using clusters~\citep{Rana,Gupta}, it is complementary to the techniques and datasets used in the above works and invokes less number of assumptions. 

Upcoming X-ray missions such as eRosita~\citep{erosita} (to be launched next year in 2019) and Athena~\citep{Athena} should be able to improve upon the limits set in this paper. However, detailed forecasts will be considered in future works.

\begin{acknowledgements}
Sajal Gupta is  supported by a DST-INSPIRE fellowship. \rthis{We would like to thank the anonymous referees for detailed critical feedback on the manuscript, which forced us to revisit some of our assumptions}. We are grateful to Alexey Vikhlinin for providing us with the data from V06 and useful correspondence and also to I-Non Chiu for helpful discussions.
 \end{acknowledgements}
\bibliography{main}
\end{document}